\documentstyle[preprint,tighten,prd,aps]{revtex}
\begin{document}
\newcommand{\dsl}{\raise.15ex\hbox{$/$}\kern-.57em\hbox{$\partial$}}
\newcommand{\asl}{\!\not\!\! A}
\newcommand{\D}{{\cal D}}
\newcommand{\ddet}{\mathop{\rm det}\nolimits}
\newcommand{\h}   {\mathop{\rm h}\nolimits}
\newcommand{\g}   {\mathop{\rm g}\nolimits}
\newcommand{\tr}  {\mathop{\rm tr}\nolimits}
\draft
%%%%%%%%%%%%%%%%%%%%%%%%%%%%%%%%%%%%%%%%%%%%%%%%%%%%%%%%%%%%%%%%%%%%%%
\title {Thermodynamics of relativistic fermions\\
%      ========================================
             with Chern-Simons coupling}
%            ===========================
%
%
%
\author{N. Brali\'c\thanks{nbralic@lascar.puc.cl}}
\address{Facultad de F\'\i sica,
         Pontificia Universidad Cat\'olica de Chile\\
         Casilla 306, Santiago 22, Chile}
\author{D.Cabra\thanks{On leave of absence from
                       Universidad Nacional de La Plata, Argentina}}
\address{Institut f\"ur Theoretische Physik der
         Universit\"at Heidelberg\\
         Philosophenweg 16, 69120 Heidelberg, Germany}
\author{F.A. Schaposnik\thanks{Investigador CICBA}}
\address{Departamento de F\'\i sica,
         Universidad Nacional de La Plata\\
         C.C. 67, (1900) La Plata, Argentina}

\date{May, 1994}
\maketitle
\begin{abstract}
We study the thermodynamics of the relativistic Quantum Field Theory
of massive fermions in three space-time dimensions coupled to an
Abelian Maxwell-Chern-Simons gauge field.  We evaluate the specific
heat at finite temperature and density and find that the variation
with the statistical angle is consistent with the non-relativistic
ideas on generalized statistics.
\end{abstract}
%%%%%%%%%%%%%%%%%%%%%%%%%%%%%%%%%%%%%%%%%%%%%%%%%%%%%%%%%%%%%%%%%%%%%%
\narrowtext
\section{Introduction}
%       ==============
%
Statistics plays a central role in physics, and marks one of the
main parting points between the classical and quantum domains.
In conventional phenomena, taking place in $3+1$-dimensions, the
invariance under the interchange of identical particles forces the
wave function to be either symmetric or antisymmetric.  The
spin-statistics theorem in relativistic quantum field theory then
ties this symmetry to the spin, and all fundamental excitations are
classified as either fermions or bosons.  Through the spin the
fermionic or bosonic character can manifest itself even in one- or
few-particle systems.  However, it is in the thermodynamics of a
system that a symmetric or antisymmetric wave function has a
dramatic effect.

In $2+1$-dimensional phenomena new possibilities open up, and
excitations of generalized statistics interpolating between fermions
and bosons may occur.  The role of these {\it anyons\/}~%
\cite{ReviewAnyons} in the quantum Hall effect is fairly well
established~\cite{QuantumHall}, and they may be relevant also in the
superconductivity of some materials at high temperature~\cite{HighTc}.
Of course, in these and other possible phenomenological applications
in two-dimensional condensed matter systems, anyons are just low
energy descriptions of the effective behaviour of particles and
interactions which, although restricted to two-dimensions, exist in
three-dimensional space and conform to the conventional boson-fermion
classification.

However, the same reasons which make anyons possible at low energies,
pose a challenging question.  It is certainly of great theoretical
interest to establish the extent to which anyonic excitations can
exist as fundamental objects in a truly $2+1$-dimensional world, just
as ordinary fermions or bosons, or if on the contrary, they are
always a low energy illusion.  Mathematically, the possibility of
generalized statistics has its origin in the topology of the
configuration space of a many particle system in $2+1$-dimensions.
Under the double interchange of two identical particles,
$x_1-x_2 \to x_2-x_1 \to x_1-x_2$,  the observability of the
probability density requires that the wave function transforms as
$\psi \to e^{i2\theta}\psi$.  In spacetime dimensions greater than
three this double interchange is just the identity operation, so we
are forced to $\theta = 0$ (bosons) or $\theta = \pi$ (fermions).
In $2+1$-dimensions however, the winding of the trajectories followed
by the particles as they are interchanged may be non trivial, in
which case the operation is not connected to the identity and
$\theta$ remains unrestricted.

This, however, is only a mathematical possibility.  Its realization
in a quantum theory depends on non trivial issues regarding the short
distance behavior of the theory.  Indeed, the very notion of the
linking of two trajectories assumes some kind of hard core repulsive
interaction between the particles (accounting for some generalized
exclusion principle), and this may or may not be consistent with a
formulation from first principles.  In a non relativistic treatment,
the transformation law of the anyonic wave function under particle
interchange provides a well defined meaning for the concept of
generalized statistics.  In practice this is implemented by minimally
coupling the matter fields to an Abelian Chern-Simons gauge field.
However, the non existence of an ideal anyon gas, analogous to the Bose
and the Fermi gases, represents a major obstacle in the understanding of
anyonic thermodynamics.  A major tool in this respect has been the
perturbative expansion in the statistical angle $\theta$, which when
done at the bosonic end exhibits the singular nature of the
statistical interaction at short distances~\cite{ThetaExp}.  Although
at the computational level this can be regularized with a repulsive
$\delta$-function contact potential, these short distance difficulties
are beyond the scope of the non relativistic theory.

The physics of anyons at short distances must be formulated and
discussed in the framework of relativistic quantum field theory.
This has been done mostly in the canonical formalism~\cite{Canonical},
where the main difficulty is again the non existence of free
asymptotic many particle states.  Besides posing technical
difficulties which are still controversial~\cite{Canonical}, this
obscures somewhat the concept of generalized statistics in this
framework.  To a large extent, it has acquired in the literature a
rather algebraic meaning, in terms of the phase factor which enters
generalized canonical commutation relations, again interpolating
between the conventional commutator and the anticommutator for
bosons and fermions.  But those algebraic relations are rather
formal, so long as there is no direct relation with free asymptotic
states, and they do not address the relevant short distant issues,
so they may or may not turn out to be consistent with a proper
regularization procedure.

Since one expects that the statistics obeyed by the fundamental
excitations will have direct consequences in the thermodynamic
properties, a possible way around these difficulties is to use
the functional integral formulation to study the theory at finite
temperature, thus avoiding such difficult issues as the lack of free
asymptotic states or the exact kinematical meaning of generalized
statistics in the relativistic theory.  Here we adopt that point of
view, and report our first results for some thermodynamic properties
of massive fermions in three space-time dimensions coupled to an
Abelian Maxwell-Chern-Simons field.  In particular, we evaluate the
specific heat at finite temperature and density for different values of
the statistical angle $\theta$.  The calculation is done to leading order
in the conventional loop expansion, complemented with a low momentum
approximation which allows for further simplifications.

We should admit from the start a limitation of our approach.  At
least within the loop expansion, the theory is not properly defined
unless the Maxwell term is included in the pure gauge action.  (Note,
however, that a Maxwell term is generated anyhow by the quantum
fluctuations of the matter fields, even if absent at the classical
level).  As it happens, that term is also expected to conceal the
effect of the generalized statistic induced by the Chern-Simons term,
at least at short distances~\cite{MaxTerm}.  Thus, we do not expect to
see in our results a very notorious shift, say, from fermionic to
bosonic properties as we vary the statistical angle $\theta$.  To
complicate matters further, in $2+1$-dimensions there is no
Bose-Einstein condensation, so the specific heat does not exhibit
a dramatic difference between fermions and bosons, as we are used
to in $3+1$-dimensions.  Nevertheless, our results do show that
the dependence of the specific heat in the statistical angle is
consistent with a smooth transit from a fermionic to a bosonic
behavior.
%%%%%%%%%%%%%%%%%%%%%%%%%%%%%%%%%%%%%%%%%%%%%%%%%%%%%%%%%%%%%%%%%%%%%%
\section{The theory at zero chemical potential}
%       =======================================
\label{sec:MuZero}

We consider massive fermions in three Euclidean dimensions coupled
to an Abelian Maxwell-Chern-Simons gauge field with a Lagrangian
\begin{equation}
{\cal L} =
    \bar{\psi} (i\dsl + m) \psi
 +  \frac{1}{4} F^2_{\mu\nu}
 - i\frac{e^2}{8\theta} \epsilon_{\mu\nu\lambda} F_{\mu\nu} A_\lambda
 + e\bar{\psi} \asl \psi,
\label{Lag0}
\end{equation}
where $e^2$ has dimensions of mass and the statistical angle $\theta$
is a dimensionless parameter.  Our choice of $\gamma$-matrices is
$\gamma_{i}=i\sigma_i$, with $\sigma_{i}$ the usual Pauli matrices.

In order to establish conventions, describe our treatment and review
previous results, we shall evaluate first the partition function at
finite temperature and zero chemical potential.  In the following
section we introduce a non-vanishing chemical potential to discuss
the thermodynamics of the system at finite fermion density.

We carry out the finite temperature calculations as usual,
compactifying the (Euclidean) time variable into the range
$0 \le \tau \le \beta = 1/T$ (in our units, $\hbar = c = k = 1$).
Then, the functional integral defining the partition function should
be computed using periodic (antiperiodic) boundary conditions (in
time) for bosons (fermions).  The partition function is then defined
as
\begin{equation}
{\cal Z} =
  {\cal N}(\beta) \int \D\bar{\psi} \D\psi \D A_\mu \>
                  \exp\left( -S[A_\mu] \right) ,
\end{equation}
with
\begin{equation}
S[A_\mu] =
         \int_0^{\beta} d\tau \int d^2x \, {\cal L}
         \equiv \int_\beta d^3x \, {\cal L} .
\end{equation}
The relevance of the normalization factor ${\cal N}(\beta)$ will be
discussed below.  Integrating out the fermions one has
\begin{equation}
{\cal Z} =
  {\cal N}(\beta) \ddet_{(-)}(i\dsl + m)
   \int \D A_\mu \>
   \exp\left[ -\left( S[A_\mu] + S_q[A_\mu] \right) \right] ,
\end{equation}
where the subindex $(-)$ means that the determinant has to be
evaluated using antiperiodic boundary conditions, and
\begin{equation}
S_q[A_\mu] =
  -\ln \ddet_{(-)} \left( 1 + \frac{e\asl}{i\dsl + m } \right)
\end{equation}
is the contribution of the fermionic quantum fluctuations to the
effective action.  To one loop order this is
\begin{equation}
S_q[A_\mu] =
  \frac{e^2}{2} \tr \left( \frac{\asl}{i\dsl + m } \right)^2 .
\label{Sq1}
\end{equation}
When computing the trace, the antiperiodic boundary conditions are
implemented in momentum space replacing integrals over $p^0$ by sums
over the discrete Matsubara frequencies
\begin{equation}
p^0_n = \frac{(2n+1)\pi}{\beta}
\label{Matsubara0}
\end{equation}

The one-loop result for $S_q$ in Eq.\ (\ref{Sq1}) can be evaluated
in closed form~\cite{Aitchinson}.  However, to simplify the numerical
analysis, here we shall work to leading order in an expansion in
powers of the momenta.  Then, following Ref.~\cite{DP} one finds
\begin{equation}
S_q[A_\mu] =
  \h(\beta) \int_\beta d^3x
  \left( \frac{1}{6} F^2_{\mu\nu} -
  \frac{im}{2} \epsilon_{\mu\nu\lambda} F_{\mu\nu} A_\lambda \right) ,
\label{Sq0}
\end{equation}
where the function $\h(\beta)$ is given by
\begin{equation}
\h(\beta) = \frac{e^2}{8m\pi} \tanh (m\beta /2) .
\end{equation}
Then, in this approximation the partition function is given by a
Gaussian integral over the gauge field, times the partition function
of free massive fermions:
\[
{\cal Z} =
  {\cal N}(\beta) \ddet_{(-)}(i\dsl + m) \int \D A_\mu \>
  \exp\left( -S_{eff}[A_\mu] \right) ,
\]
where the effective action $S_{eff} = S + S_q$ is given by
\begin{equation}
S_{eff}[A_\mu] =
    \frac{1}{4} \int_\beta d^3x \,
    \left\{
    \left(1 + \frac{2}{3} \h(\beta) \right) F^2_{\mu\nu}
  -i\left(\frac{e^2}{2\theta} + 2m \h(\beta) \right)
    \epsilon_{\mu\nu\lambda} F_{\mu\nu} A_\lambda
    \right\} .
\end{equation}
{}From the coefficient of the Maxwell term, one sees that, as usual,
the loop expansion requires $\alpha = e^2/4\pi m \ll 1$.  Similarly,
from the coefficient of the Chern-Simons term one would conclude that
in this case the loop expansion also requires $\theta \ll 2\pi$.  Yet,
from the experience gained with  three-dimensional fermions at zero
temperature~\cite{Coleman,GMSS}, one could hope that this
restriction over $\theta$ may not really apply.  This, however, is
controversial at finite temperature~\cite{SemPopKao}.

As discussed in~\cite{Bernard}, at finite temperature one has to take
into account the contribution from the Faddeev-Popov determinant
arising from gauge fixing, even if the theory is Abelian.  We shall
choose the Lorentz gauge ($\partial_\mu A_\mu = 0$) for which the
Fadeev-Popov determinant is given by
\begin{equation}
\Delta_{FP} = \ddet_{(+)} (-\partial^2)
\end{equation}
where now the  $(+)$  subindex indicates that the determinant has
to be evaluated using periodic boundary conditions~\cite{Bernard}.
Evaluating the contribution to the determinants coming from the
space-time indices and keeping track of all temperature dependent
factors, one obtains
\begin{equation}
{\cal Z} =
  \ddet_R (i\dsl + m) \times \ddet_R^{-1/2}(-\partial^2 + M^2) ,
\label{PartFuncI3}
\end{equation}
where $M^2$ is defined by
\begin{equation}
M^2 =
  \left[
  \frac {e^2/\theta + 4 m \h(\beta)} {2 + (4/3)\h(\beta)}
  \right]^2 ,
\end{equation}
and the subindex $R$ denotes that one must take the finite part of
the determinants: temperature dependent divergences coming from these
determinants and from the normalization factor ${\cal N}(\beta)$
cancel each other~\cite{Bernard}.  Also, following~\cite{Actor}, in
Eq.\ (\ref{PartFuncI3}) we have omitted an overall divergent factor
which does not contribute to the thermodynamic properties of the
system.

It is interesting to note that the partition function (\ref{PartFuncI3})
corresponds to an effective theory with a free massive fermion and a
free boson with a mass $M$ which is related to the Chern-Simons
topological mass.  This mass acquires a temperature dependence which
arises from the fermion determinant.  It should be stressed that all
dependence on the statistical angle $\theta$ occurs through that mass.

Determinants are easily evaluated.  Standard finite temperature
methods~\cite{Bernard} lead to
\begin{eqnarray}
\ln \ddet_{R}(i\dsl + m) & = &
      V \int \frac{d^2p}{(2\pi)^2}
        \ln( 1 + e^{-\beta \sqrt{p^2 + m^2} } )
\nonumber\\
\ln \ddet_{R}(-\partial^2 + M^2) & = &
    2 V \int \frac{d^2p}{(2\pi)^2}
        \ln( 1 - e^{-\beta \sqrt{p^2 + M^2} } )
\nonumber\\
& & +2 \beta V \Delta\Omega_R  ,
\end{eqnarray}
where  $\Delta\Omega_R$  is the finite contribution to the bosonic
determinant coming from the ``zero point energy'':
\begin{equation}
\left. \Delta\Omega_R \equiv
    \frac{1}{2} \int \frac{d^2p}{(2\pi)^2}
    \sqrt{p^2 + M^2} \right\vert_R
  = - \frac{1}{12\pi} M^3 .
\end{equation}

The thermodynamic potential is defined as
\begin{equation}
\Omega = -\frac{1}{\beta V} \ln {\cal Z} ,
\end{equation}
so that in the present case we have
\begin{equation}
\Omega =
  - \frac{2}{\beta} \int \frac{d^2p}{(2\pi)^2}
    \ln (1 + e^{-\beta \sqrt{p^2 + m^2} })
  + \frac{1}{\beta} \int \frac{d^2p}{(2\pi)^2}
     \ln (1 - e^{-\beta \sqrt{p^2 + M^2} }) .
\end{equation}
It should be stressed that we have not included in the thermodynamic
potential the zero-point energy contribution  $\Delta \Omega_R$, which
is eliminated by an appropriate normal ordering~\cite{HaberWeldon}.

Now, using $C_v = -\beta^2\, \partial^2(\beta\Omega)/\partial\beta^2$,
we get
\widetext
\begin{eqnarray}
C_v &=&
    \frac{1}{4\pi\beta^2} \int_{\beta m}^{\infty} dx
  \,\frac{x^3}{\cosh^2(x/2)} +
    \frac{1}{8\pi\beta^2} \int_{\beta M}^{\infty} dx
  \,\frac{x^3}{\sinh^2(x/2)}
\nonumber \\
& &+
    \frac{1}{\pi} \frac{\partial M}{\partial \beta}
    \left( \frac{\beta^2 M^2} {\exp(\beta M) - 1} \right) +
    \frac{1}{2\pi} \frac{\partial^2 M}{\partial \beta^2}
    \left( \beta^2 M \ln (1 - \exp(-\beta M) ) \right)
\nonumber \\
& &+
    \frac{1}{2\pi} \left( \frac{\partial M}{\partial \beta} \right)^2
    \left( -\beta^3 M + \beta^2 \ln (1 - exp(-\beta M) ) -
    \frac{\beta^3 M} {\exp(-\beta M) - 1} \right) .
\end{eqnarray}
\narrowtext
Except for being at zero chemical potential, this result was the goal
of our calculation.  It is not usual in field theory to look at
purely thermodynamic functions like $C_v$.  But, as this calculation
has illustrated, for our purposes is important since it gives direct
information about the quantum excitations of the system, without need
of facing difficult issues related to the asymptotic states.  In the
next section we extend this calculation to incorporate a finite
fermionic density.
%%%%%%%%%%%%%%%%%%%%%%%%%%%%%%%%%%%%%%%%%%%%%%%%%%%%%%%%%%%%%%%%%%%%%%
\section{The theory at non-zero chemical potential}
%       ===========================================
%
To describe the system at finite fermion density we introduce a
chemical potential $\mu$, so the Lagrangian in Eq.\ (\ref{Lag0})
becomes
\begin{equation}
{\cal L} =
     \bar{\psi} (i\dsl + m -i\gamma_0 \mu) \psi
   + \frac{1}{4} F^2_{\mu\nu}
  -i \frac{e^2}{8\theta} \epsilon_{\mu\nu\lambda} F_{\mu\nu} A_\lambda +
   e \bar{\psi} \asl \psi .
\end{equation}
The contribution of the fermionic quantum excitations to the effective
action is now
\begin{equation}
S_q[A_\mu] =
  - \ln \ddet_{(-)}
    \left ( 1 + \frac{e\asl} {i\dsl + m - i\gamma_0\mu} \right)
\end{equation}
Comparing this with Eq.\ (\ref{Sq0}) we see that the presence of the
chemical potential $\mu$ amounts to a shift of the Matsubara
frequencies  $p^0_n$  in Eq.\ (\ref{Matsubara0}) to
\begin{equation}
p^0_n = \frac{(2n+1)\pi} {\beta} - i\mu ,
\end{equation}
Then, repeating the previous calculation with these shifted frequencies,
a tedious but straightforward calculation leads to the effective action
\begin{equation}
S_{eff}[A_\mu] =
  \frac{1}{4}  \int_\beta d^3x
    \left\{
    \left(1 + \frac{2}{3} \g(\beta,\mu) \right) F^2_{\mu\nu}
 -i \left(\frac{e^2}{2\theta} + 2m \g(\beta,\mu) \right)
    \epsilon_{\mu\nu\lambda} F_{\mu\nu} A_\lambda
    \right\}
\end{equation}
where now
\begin{equation}
\g(\beta,\mu) =
  \frac{e^2}{16\pi m}
  \left[
  \tanh\left( \frac{\beta}{2} (m - \mu) \right)
 +\tanh\left( \frac{\beta}{2} (m + \mu) \right)
  \right] .
\end{equation}
This result holds, again, to leading order in a low momentum expansion.
Moreover, we have neglected in this expression Lorentz non-invariant
terms which are of order  $\beta \exp(-\beta m)$.  These terms are
negligible for  $\beta m$  sufficiently big, which is anyway assumed
by the low momentum approximation.

Following the same steps as in Sec.\ \ref{sec:MuZero}, for the
thermodynamic potential we obtain
\begin{eqnarray}
\Omega &=&
  -\frac{1}{\beta} \int \frac{d^2p}{(2\pi)^2}
   \ln (1 + e^{-\beta(\sqrt{p^2 + m^2} + \mu)} \,)
  -\frac{1}{\beta} \int \frac{d^2p}{(2\pi)^2}
   \ln (1 + e^{-\beta(\sqrt{p^2 + m^2} - \mu)} \,)
\nonumber\\
& &
 + \frac{1}{\beta}\int \frac{d^2p}{(2\pi)^2}
   \ln (1 - e^{-\beta\sqrt{p^2 + M_{(\mu)}^2}} \,) ,
\end{eqnarray}
where now
\begin{equation}
M_{(\mu)}^2 =
  \left[
  \frac{e^2/\theta + 4m\g(\beta,\mu)}
       {2 + (4/3)\g(\beta,\mu)}
  \right]^2  .
\end{equation}
For the specific heat we now find
\widetext
\begin{eqnarray}
C_v &= &
  \frac{1}{8\pi\beta^2}
  \left[
  \int\limits_{\beta(m+\mu)}^{\infty} dx\>
    \frac{(x - \beta\mu)x^2}{\cosh^2(x/2)}
+ \int\limits_{\beta(m-\mu)}^{\infty} dx\>
    \frac{(x + \beta\mu)x^2}{\cosh^2(x/2)}
+ \int\limits_{\beta M_{(\mu)}}^{\infty} dx\>
    \frac{x^3}{\sinh^2(x/2)}
  \right]
\nonumber \\
& &
+ \frac{1}{\pi}\frac{\partial M_{(\mu)}}{\partial\beta}
  \left( \frac{\beta^2 M^2_{(\mu)}}{\exp(\beta M_{(\mu)}) - 1} \right)
+ \frac{1}{2\pi} \frac{\partial^2 M_{(\mu)}}{\partial \beta^2}
  \left[ \beta^2 M_{(\mu)}\ln (1 - \exp(-\beta M_{(\mu)})) \right]
\nonumber \\
& &
+ \frac{1}{2\pi}
  \left( \frac{\partial M_{(\mu)}}{\partial\beta} \right)^2
  \left(
 -\beta^3 M_{(\mu)} + \beta^2 \ln(1 - \exp(-\beta M_{(\mu)})) -
  \frac{\beta^3 M_{(\mu)}} {exp(-\beta M_{(\mu)}) - 1}
  \right) .
\label{CvMu}
\end{eqnarray}
\narrowtext
Similarly, for the fermion number density
\mbox{$\rho = -\partial\Omega/\partial \mu$}, we get
\begin{eqnarray}
\rho &=&
  \int\frac{d^2p}{(2\pi)^2}
  \left[
  \frac{1}{1 + e^{\beta(\sqrt{p^2 + m^2} + \mu)}} -
  \frac{1}{1 + e^{\beta(\sqrt{p^2 + m^2} - \mu)}}
  \right]
\nonumber\\
& &-
  \frac{1}{2\pi\beta} M_{(\mu)}
  \frac{\partial M_{(\mu)}}{\partial\mu} \ln
  \left[
  2 \sinh \left( \frac{\beta}{2} M_{(\mu)} \right)
  \right]
\label{rho}
\end{eqnarray}
This is a constraint equation which must be inverted to determine the
chemical potential $\mu$ in terms of a given density and temperature.
It should be noticed that the first two terms in this expression
correspond to the free fermion (relativistic) gas, with both fermions and
anti-fermions contributing to the density.  Similarly, in Eq.\ (\ref{CvMu}),
the first three terms correspond to the contributions to $C_v$ from free
fermions and anti-fermions of mass $m$ and chemical potential $\pm\mu$,
and from free bosons of mass $M_{(\mu)}$ without a chemical potential.
The numerical analysis of these equations is the subject of the next
section.
%
%%%%%%%%%%%%%%%%%%%%%%%%%%%%%%%%%%%%%%%%%%%%%%%%%%%%%%%%%%%%%%%%%%%%%%
\section{Numerical results}
%       ===================
%
In this section we present our numerical results.  For given values
of the statistical angle $\theta$ and the fermionic density $\rho$,
we invert Eq.\ (\ref{rho}) numerically (with a precission of one part
in $10^5$), and then substitute the chemical potential into
Eq.\ (\ref{CvMu}) to obtain the specific heat $C_v$ as a function of
the temperature at different values of $\rho$ and $\theta$.  Since
here both $e^2$ and $m$ have dimesions of mass, we find it useful to
parametrize the theory in terms of the dimensionless variables
$\alpha = e^2/4\pi m$, $\rho / m^2$, $\beta m$ and $\theta$.

We have performed numerical computations covering the following domain
of parameters:
\begin{eqnarray}
\alpha           &\in& (10^{-2}, 10^{-6})
\nonumber\\
\frac{\rho}{m^2} &\in& (10^{-8}, 10)
\nonumber\\
\theta           &\in& (10^{-4}, 1)
\end{eqnarray}
and in the temperature range
\begin{equation}
\beta m           \in (1, 10^2)
\end{equation}
There are no qualitative differences for the results over the whole
domain of parameters.  Typical curves for the specific heat as a function
of the temperature are presented in Figs.\ \ref{fig1} and \ref{fig2} for
$\theta = 1$ and $\theta = 10^{-4}$, respectively, at
$\alpha=9.73\times 10^{-7}$, and $\rho/m^2=1.78\times 10^{-8}$.  These
are values relevant for high $T_c$  superconductors~\cite{Hosotani}, and
correspond to $m = 7.5\times 10^{10}\,\text{cm}^{-1}$, thrice the value of
the bare electron mass, and a fine structure constant in $2+1$ space-time
dimensions obtained from that in $3+1$ space-time dimensions dividing
by the interplanar spacing.

Although the curves in Figs.\ \ref{fig1} and \ref{fig2} seem almost
identical, the specific heat in Fig.\ \ref{fig1} ($\theta = 1$) is
slightly larger than that in Fig.\ \ref{fig2} ($\theta = 10^{-4}$).
This is a characteristic feature, which we find to hold in the entire
parameter range, although the actual difference depends on the values
of $\alpha$ and $\rho$.  We think it is a very interesting result,
since it is consistent with the (unimpressive) difference between the
free bosonic and fermionic specific heats in $2+1$-dimensions.  This
is illustrated in more detail in Figs.\ \ref{fig3} and \ref{fig4}.
In Fig.\ \ref{fig3} we present the difference of the specific heats
$C_v(\text{bosonic}) - C_v(\text{fermionic})$ for the free relativistic
(charged) boson/fermion gases, both at $\rho/m^2 = 0.01$.  This is to be
compared with Fig.\ \ref{fig4}, where we present the difference
$C_v(\theta=1) - C_v(\theta = 10^{-4})$, both at $\alpha = 0.01$ and
$\rho/m^2 = 0.01$.  Although this is not a proof of a shift from fermionic
towards bosonic statistics (whatever that means in this relativistic
system), it is certainly consistent with that intuitive picture: since we
start from a fermionic system at $\theta = 0$, one expects a transition
towards a bosonic behavior as $\theta$ is increased, provided the
nonrelativistic ideas about Chern-Simons-generalized statistics hold in
this case.

In connection with the curves shown in Figs.\ \ref{fig1}--\ref{fig4},
it is worthwhile noticing that their apparent violation of the classical
Dulong-Petit law is not so.  For a $2+1$-dimensional non-relativistic
classical ideal gas one finds $C_v = \rho$.  In our case, however, $\rho$
stands for the Abelian charge density, and therefore is the density of
particles minus the density of anti particles, as shown by the first
terms in Eq.\ (\ref{rho}) in the case of fermions.  Hence, one can indeed
reach large values of $C_v$ without a corresponding increment of $\rho$.

We have studied also the fermion density as a function of the chemical
potencial.  We obtained a smooth function which does not exhibit the
step structure found in~\cite{Son} where it is interpreted as a signal
of a critical behavior.  Note, however, that that reference considers
a high temperature range (opposite from ours) and starts from a pure
Chern-Simons system, without a Maxwell term.
\section{Conclusions}
%       =============
%
We have analyzed at finite temperature and density a system of massive
fermions in three space-time dimensions coupled to an Abelian
Maxwell-Chern-Simons field.  In particular, we evaluated the specific
heat for different values of the statistical angle $\theta$.  This was
done to leading order in the loop expansion and in the low momentum
approximation.  In doing so, we have shown that it is possible to obtain
physical information of the system without having to give an explicit
answer to such difficult issues as the lack of free asymptotic states or
the exact kinematical meaning of generalized statistic in relativistic
quantum field theory.

Our main result, the specific heat as a function of temperature and
density (Eqs.\ (\ref{CvMu}-\ref{rho})), was analysed in a wide range of
parameters within the low-temperature regime.  This limitation to low
temperatures arises in part from the low momentum expansion, which we
here adopted for simplicity but can be avoided \cite{Aitchinson}.
However, to move to higher temperatures one will also have to face the
contribution of Lorentz non-invariant terms which here we have discarded.
We hope to come back to this point in a future work.

As we ourselves had expected,  the specific heat has only a mild
dependence on the statistical angle $\theta$. However, this dependence
is consistent with what one expects from the nonrelativistic picture of
generalized statistics, thus suggesting that those ideas may still be
applicable, to some extent, in the relativistic domain.

As we pointed out earlier, a more marked dependence on the statistical
angle $\theta$ may be expected in a pure Chern-Simons theory without
Maxwell term.  That, however, will possibly require a non-perturbative
treatment.  In this respect it may be of interest to consider other
thermodynamic functions which may be more sensible to the long-distance
properties, and therefore more immune to the presence of the Maxwell term.
\section*{Acknowledgments}
This work was partially supported by FONDECYT, under grant 751/92, by CONICET,
and by Fundaci\'on Andes and Fundaci\'on Antorchas, under grant 12345/9.
%%%%%%%%%%%%%%%%%%%%%%%%%%%%%% REFERENCES

%%%%%%%%%%%%%%%%%%%%%%%%%%%%%%  FIGURES
%\section*{Figure Captions}
%        =================

\begin{figure}
\caption{Specific heat as a function of temperature at
  $\alpha = 9.73 \times 10^{-7}$, $\rho /m^2 = 1.78 \times 10^{-8}$
  and $\theta = 1$.}
\label{fig1}
\end{figure}

\begin{figure}
\caption{Specific heat as a function of temperature at
  $\alpha = 9.73 \times 10^{-7}$, $\rho /m^2 = 1.78 \times 10^{-8}$
  and $\theta = 10^{-4}$.}
\label{fig2}
\end{figure}

\begin{figure}
\caption{Difference of the specific heats $C_v(\mbox{bosonic})$ and
  $C_v(\mbox{fermionic})$ for the free relativistic
  (charged) boson/fermion gases, both at $\rho/m^2 = 0.01$.}
\label{fig3}
\end{figure}

\begin{figure}
\caption{Difference of the specific heats $C_v(\theta=1)$ and
  $C_v(\theta=10^{-4})$, both at $\alpha=0.01$ and $\rho/m^2=0.01$.}
\label{fig4}
\end{figure}

%%%%%%%%%%%%%%%%%%%%%%%%%%%%%%%%%%%%%%%%%%%%%%%%%%%%%%%%%%%%%%%%%%%%%%
\end{document}